# Advanced Non-Destructive in Situ Characterization of Metals with the French Collaborating Research Group D2AM/BM02 Beamline at the European Synchrotron Radiation Facility

Gilbert André Chahine [1,*], Nils Blanc [2], Stephan Arnaud [2], Frédéric De Geuser [1], René Guinebretière [3] and Nathalie Boudet [2]

[1] SIMaP, Grenoble INP, CNRS, Université Grenoble Alpes, 38000 Grenoble, France; Frederic.De-Geuser@simap.grenoble-inp.fr
[2] Institut Néel, CNRS, Université Grenoble Alpes, 38000 Grenoble, France; nils.blanc@ neel.cnrs.fr (Ni.B.); stephan.arnaud@neel.cnrs.fr (S.A.); nathalie.boudet@ neel.cnrs.fr (Na.B.)
[3] IRCER, UMR 7315, CNRS, Centre Européen de la Céramique, Université de Limoges, F-87068 Limoges, France; rene.guinebretiere@unilim.fr
* Correspondence: gilbert.chahine@esrf.fr; Tel.: +33-476-881-778

**Abstract:** The ability to non-destructively measure the structural properties of devices, either in situ or operando, are now possible using an intense X-ray synchrotron source combined with specialized equipment. This tool attracted researchers, in particular metallurgists, to attempt more complex and ambitious experiments aimed at answering unresolved questions in formation mechanisms, phase transitions, and magnetism complex alloys for industrial applications. In this paper, we introduce the diffraction diffusion anomale multi-longueur d'onde (D2AM) beamline, a French collaborating research group (CRG) beamline at the European Synchrotron Radiation Facility (ESRF), partially dedicated to in situ X-ray scattering experiments. The design of the beamline combined with the available equipment (two-dimensional fast photon counting detectors, sophisticated high precision kappa diffractometer, a variety of sample environments, continuous scanning for X-ray imaging, and specific software for data analysis) has made the D2AM beamline a highly efficient tool for advanced, in situ synchrotron characterization in materials science, e.g., single crystal or polycrystalline materials, powders, liquids, thin films, or epitaxial nanostructures. This paper gathers the main elements and equipment available at the beamline and shows its potential and flexibility in performing a wide variety of temporally, spatially, and energetically resolved X-ray synchrotron scattering measurements in situ.

## 1. Introduction

The race for miniaturization, motivated by both the observation of new promising properties when scaling down devices and structures and the need to understand the local phenomena in classical devices and materials, led to an unprecedented race in improving characterization techniques for local probing. New avenues for material engineering are emerging thanks to a better understanding of the local heterogeneity. This is due to our ability to follow physical and chemical phenomena taking place in these devices when exposed to external forces, reactions, or simply ageing. While the majority of these techniques still requires specific sample preparation and very restrictive environments, X-ray scattering stands out as a non-destructive measurement tool for structural parameters. Due to the penetration depth of X-rays and the brilliance of third generation synchrotron sources, X-ray scattering-based techniques have become an essential tool for the investigation of complex micro- and nano-systems even when buried in other materials. In parallel, materials science is now defining new roadmaps to optimize composites and more complex material systems when aiming for multifunctional devices. The achievement of this goal depends on our understanding of the structure–properties relationship on the micro- and nano-scale levels. The tunability of these properties is reliant on our ability to not only image the structure, but mainly to follow the response of its parameters under external loads. Therefore, in situ characterization can offer more complete fundamental insights into the structural evolutions. This will allow the understanding and designing of novel alloys to prevent damage and reduce corrosion phenomena, thus fulfilling industrial requirements in terms of performance, competitiveness, and profitability. This is why most of the synchrotron X-ray-based techniques are more highly requested, as it delivers structural sensitivity in a flexible, in situ environment.

Among these techniques, using a monochromatic X-ray beam, three-dimensional reciprocal space mapping (3D-RSM) can be performed either close to the origin of the reciprocal space or further away through the use of appropriate rotations of both the sample and the detector. Wide-angle measurements yield lattice tilt, composition, thickness, and strain distributions in single crystals as well as polycrystalline materials that can be heterogeneous bulk or powdered materials, thin films, or epitaxial nanostructures. 3D-RSM offers a clearer view of the structural evolution in polycrystalline samples, as is the case of most metallic alloys, such as Al-based alloys, where mapping local strain distributions have shown opposite behaviors in the crystal lattice under thermomechanical loading within micro-grain compared to grain boundaries [1].

Small-angle X-ray scattering (SAXS) gives access to phenomena taking place at the nanoscale, such as precipitation, diffusion, and phase separation, which affects the mechanical properties of metallic alloys. The microstructural evolution in terms of nanoparticles size, shape, and volume fraction during aging can be mapped using in situ SAXS [2]. The kinetics of precipitates may consequently be resolved in heterogeneous samples by following, for example, element contents across a diffusion weld, [3].

Moreover, grazing incidence (GI) measurements can be performed to investigate thin layers or surfaces such as the study of the evolution of stepped vicinal surfaces during thermal treatment [4].

The use of synchrotron X-ray techniques offers, additionally, the possibility to tune the X-ray wavelength. This feature represents a powerful tool for heterogeneous systems having multiple scattering contributions in the same reciprocal space region. Subsequently, to avoid overlapping peaks, it is possible to extinguish one of the peaks by tuning the energy of the incident beam at the absorption edge of that element. Based on the variation of the atomic scattering factor near the absorption edges of a given element, this element-specific probe tool (anomalous scattering) can be coupled with XRD, SAXS, and other experimental configurations. The combination of anomalous scattering and SAXS allowed, for instance, the determination of volume fraction and chemical composition in alloys [5,6]. Anomalous small-angle X-ray scattering in a grazing incidence geometry (AGISAXS) has been employed to track the influence of phase separation on self-organization in multi-metallic oxide films [7].

While such X-ray-based experiments can be performed on separate synchrotron beamlines at different times, only a few of them offer the possibility of combining and coupling these tools. The diffraction diffusion anomale multi-longueur d'onde (D2AM) beamline at the European Synchrotron Radiation Facility (ESRF) has been built and designed to combine XRD, wide-angle X-ray scattering (WAXS), SAXS, GI, and anomalous scattering for materials in their native (or operando) state. In situ simultaneous SAXS/WAXS measurements can be performed at the D2AM beamline, where a specific detector was developed and used for this purpose [8]. This approach offers unique insight into the structural behavior under applied external fields. While the WAXS signal allows the scientist to follow phase transitions by tracking Bragg peak shifts or intensity fluctuations, the SAXS signal provides information about particle sizes, shapes, and phase anisotropy. All these features yield a suite of tools that may be combined/coupled to meet the users' requirements.

This paper gives an overview of the D2AM beamline, aiming to show its capabilities with respect to the characterization of metallic materials, and it offers a detailed instrumental and technical database for researchers to plan experiments. A detailed description of the beamline from the source to the sample will be described, followed by the main equipment used at the D2AM beamline and its specifications. The last part is dedicated to the sample environments frequently used at the beamline for in situ (or operando) measurements and their specifications.

## 2. Description of the Beamline

### 2.1. Primary Optics

The D2AM beamline, a French CRG (Collaborating Research Groups) beamline, was built in the 1990s on the BM02 port at the ESRF in Grenoble, France. A polychromatic X-ray beam is produced when the accelerated electrons travel through a bending magnet of 0.8 T. It is characterized by a vertical divergence of 0.2 mrad. Whereas, the effective horizontal divergence is determined by the projection of the curvature of the bending magnet, and is equivalent to 3 mrad. Due to this intrinsic and geometrical divergence, and in order to collect the maximum flux on the sample, the X-ray beam is focused back onto the sample by the X-ray optical elements mounted in the optics cabin.

Figure 1 displays the beam trajectory as well as the optical elements installed in the beamline optics cabin used for collimating and directing the X-rays towards the sample. An initial water-cooled mirror of 1.1 m length, mounted in the horizontal plane 26 m away from the source, allows a vertical reflection of the incident polychromatic beam. The surface of the mirror is coated with two strips of 400 Å-thick and 30 mm-wide Rh and Pt along the whole length of the mirror. The incidence angle of the first mirror is set slightly below the critical angle of total reflection for a wavelength $\lambda$ to reject its $\lambda/3$ harmonics. This rejection is necessary to prevent the overlapping of both contributions.

Being fixed to two benders, the mirror can be cylindrically bent, depending on the required energy, in order to parallelize the reflected beam. Hereafter, the beam is monochromatized, at an energy ranging from 5 to 40 keV, by being diffracted from an initial Si (111) water-cooled crystal mounted 1.93 m downstream of the first mirror. The monochromatic beam then hits a second Si (111) crystal, mounted on a specific sophisticated mechanical system, leading to an energy resolution $\Delta\lambda/\lambda = 2 \times 10^{-4}$. The latter is designed to allow not only tilt and position adjustment of the second crystal but also its bending in order to focus the beam sagittally. To avoid anticlastic effects while bending [9,10], the second crystal of the monochromator was locally backside-etched to shape segments of 700 µm-thick Si, separated by non-etched ribs with 400 µm widths and 3 mm thicknesses. A rib-free central zone 5 mm in width is mainly used to deliver a uniform, focused beam without the satellites that may be reflected by the inter-rib regions. The tuning of the second crystal is permanently monitored and adjusted using a feedback tracking system where a piezo motor adjusts the Bragg angle of the crystal to maintain the highest diffracted intensity when scanning the energy. Having the rotation axis of the monochromator system set on the surface of the first crystal, together with the available translations of the second crystal, delivers a point-focused fixed exit beam essential for energy scans. It is therefore possible to perform anomalous X-ray scattering by maintaining the illumination on the same spot on the sample. After the second crystal, the horizontally focused monochromatic beam gets reflected, 1.82 m later, downwards by a second mirror identical to the first one. This reflection, coupled with a rotation of the mirror around a horizontal axis perpendicular to the beam direction, puts the beam back on its initial horizontal axis defined by the source. As for the first mirror, a similar set of benders induces an ellipsoidal curvature to the second mirror to focus the beam vertically. Without using any other focusing lenses or mirrors, this optical element configuration delivers a flux of $10^{11}$ photons/s in a 100 µm × 90 µm (horizontal × vertical) focal spot size.

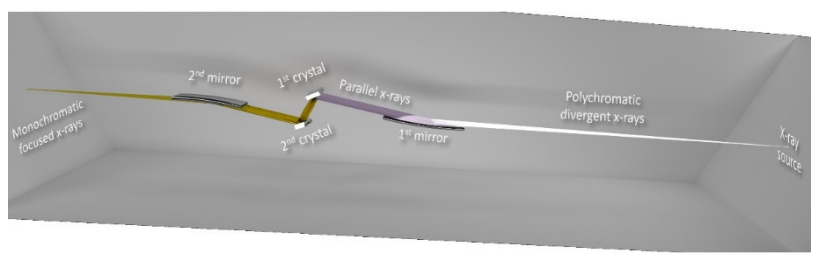

**Figure 1.** Sketch and description of the diffraction diffusion anomale multi-longueur d'onde (D2AM) beamline optical and focusing elements of the optics cabin.

After the second mirror, the beam goes through a pair of slits and travels 10 m before entering an experimental cabin 9 m in length. At the entrance of this cabin, a set of two Si Kirkpatrick–Baez (KB) mirrors are installed and coated with Ir with a roughness less than 20 Å. These mirrors can be moved into or out of the beam. By bending the KB mirrors elliptically, the first mirror (300 × 50 × 15 mm$^3$) focuses the beam vertically while the second one, with a trapezoidal shape (170 × 44 − 15 × 8 mm$^2$), focuses the beam horizontally. At 8 keV, the beam can be focused down to 30 × 30 μm$^2$ with 10$^8$ photons/s. This beam size is still higher than the predicted one because of the high roughness of the KB mirrors. A new set of mirrors will be installed in the coming years capable of focusing the X-ray beam to 10 × 10 μm$^2$, and it will get to 1 × 1 μm$^2$ with the implementation of a new X-ray source within the EBS (Extremely Brilliant Source) upgrade program of the ESRF [11]. Besides reducing the beam size on the sample, the use of the KB mirrors filters the beam vibrations that may be induced by the upstream optical elements or external works occurring during beamtime. After a pair of slits and a set of automated Al and Cu attenuation foils, the beam arrives at the sample position. The sample orientation can be dictated by a six-circle diffractometer [12].

## 2.2. Diffraction Diffusion Anomale Multi-Longueur D'onde (D2AM) Endstation

The D2AM diffractometer, illustrated in Figure 2a, is a six-circle diffractometer with a kappa geometry built by the Newport Micro-controle company (Evry, France). A list of its motors, their respective mnemonics, strokes and resolutions is provided in Table 1. Along the vertical *z*-axis, a tripod-like set of encoded motors are dedicated to adjust the horizontality of the diffractometer as well as its height with respect to the X-ray beam. In addition, two motors allow the translation of the diffractometer perpendicularly to the X-ray beam's direction as well as the rotation around the *z*-axis.

**Table 1.** List of the D2AM Diffractometer Motors and Their Respective Mnemonics, Strokes, and Resolutions.

| Motors | Mnemonic | Stroke | Resolution |
|---|---|---|---|
| *TZ1–TZ2–TZ3* | TZ | 150 mm | <1 μm |
| *TZ1–TZ2–TZ3* | RX-RY | 2° | <0.001° |
| *TY1/TY2* | TY | 130 mm | <1 μm |
| *TY1/TY2* | RZ | 2° | <0.002° |
| *Mu* | mu | −20°/200° | ±0.0002° |
| *Theta* | eta | ±190° | ±0.0002° |
| *Phi* | phi | ±180° | ±0.0002° |
| *Kappa* | kap | ±190° | ±0.0002° |
| **X sample translation** | tsx | ±5 mm | ±0.1 μm |
| **Y sample translation** | tsy | ±5 mm | ±0.1 μm |
| **Z sample translation** | tsz | 28 mm | ±0.1 μm |
| **Cradle rotations X-Y** | rox-roy | ±12° | - |
| **Sample holder translations X-Y** | tox-toy | ±12.5 mm | - |
| *Nu* | nu | −20°/200° | ±0.0002° |
| *Delta* | del | −20°/200° | ±0.0002° |
| **Theta analyzer** | *tha* | ±165° | ±0.0001° |
| ***Delta*** **analyzer** | *ttha* | ±100° | ±0.0001° |
| ***Eta*** **Analyzer** | *etaa* | 0°–100° | ±0.0001° |

The *Mu* base that can rotate around the vertical axis supports the kappa sample stage. The latter is composed of three rotational motors: *Theta*, *Kappa,* and *Phi*. On top of *Phi*, an *XYZ* translation stage is used for sample alignments using encoded motors. The *Chi* rotation is based on a combination of the three motors, *Phi*, *Kappa,* and *Theta*, and rotates the sample around the beam axis and maintains the same incidence angle. From the detection side, a motorized structure with two rotations around the vertical z-axis as well as the horizontal y-axis (*Delta*, *Nu*) gives access to a wide range of the scattering vector *Q* values. The only range-limiting parameter may be the sample environment mounted for in situ/operando experiments. All these rotations coincide in a 60 µm sphere of confusion with a maximum load of 20 kg on the sample stage and 50 kg on the detector arm.

The rotations offered by the D2AM diffractometer allow following the diffracted intensity using two different hkl modes: the 4-circles (kappa) and 6-circles (kappapsic) modes. Within these modes, an orientation matrix can be defined by referencing two Bragg reflections, thus, hkl scans can be performed or Bragg reflections with complex orientations can be accessed. Moreover, a sample holder (Figure 2b inset) made of two motorized, crossed cradles (*Rox*, *Roy*) and two translations (*tox*, *toy*) can be mounted on top of the kappa sample stage, i.e., on top of the *theta*, *chi* and *phi* motors. This configuration offers the possibility of aligning the sample surface, or diffracting planes with respect to the X-ray beam, to an accuracy of a few thousandths of a degree [13] and, thereafter, to perform 3D-RSM under a constant incidence angle (the azimuthal angle). This can be done close to a given Bragg reflection or near the origin of reciprocal space (GISAXS experiments).

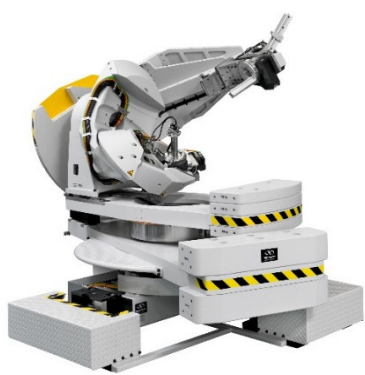

(**a**)

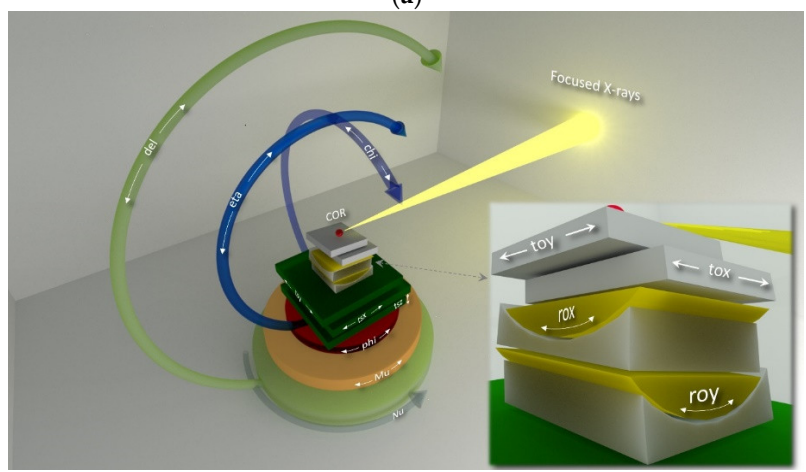

(**b**)

**Figure 2.** (**a**) The D2AM kappa diffractometer with the analyzer stage mounted on the detector arm. (**b**) The piled sample positioners, translations (*tsz*, *tsx* and *tsy*), and rotations (*Mu*, *phi*, *eta/chi*). The sample holder stage with the two translations (*tox*, *toy*) and cradles (*rox*, *roy*) are in the insert. The translations and the cradles are mounted on top of *phi*, itself, and mounted on the *eta* and *chi* rotations. All rotation axes go through the center of rotation (COR) of the diffractometer. The detector arm is connected to a vertical structure that rotates vertically along *del* and to a horizontal one that rotated along *nu* horizontally.

Finally, a 3-axis polarization analyzer system can be mounted on the detector arm. It consists of two coaxial rotations: *ThetaA* for the crystal Bragg angle and *DeltaA* for the detection angle, and a third rotation (*EtaA*) that rotates the ensemble to swap between the two polarization configurations. A large variety of analyzer crystals is available; the most commonly used ones are one flat and two curved graphite crystals and one flat germanium crystal. As the incident X-ray beam is linearly polarized in the horizontal plane, a set of polarizer crystals, with a Bragg reflection close to 90°, can be used to investigate the electronic and/or magnetic order parameters [14].

Currently, the main rotational motors of the D2AM diffractometer (*phi*, *del*, *nu*, *mu*, *kappa*, and *theta*) can be configured, together with the available detectors, to perform continuous scans. In this mode, the motor moves continuously while the detector acquires images in a synchronized manner. The scanning overhead related to the positioning and to the software/hardware communication dead time is, consequently, very small. The advantage of this approach is not only limited to a substantial gain of time (a factor of six), but it extends to offer the possibility of performing new types of exhaustive in situ measurements during very short beamtimes.

Downstream of the diffractometer, a 4 m-long granite table is fixed in the D2AM experimental cabin. Along the beam direction, this table is used for SAXS experiments where motorized sample stages and detector holders can be set up at any position along that table. It can also be used to perform GISAXS with the sample standing on the goniometer and leading to long sample–detector distances [15,16].

On the detection side, a set of three fast photon counting pixel detectors supplements the detection capabilities of the D2AM beamline. These Si-based detectors (S70, D5, and WOS [8] (WAXS Open for SAXS)) were produced by IMXPAD (Now sold by the CEGITECK INNOVATION company under the name REBIRX) and developed with the beamline staff through collaboration with CPPM (Centre de Physique des Particules de Marseille) [17]. They are characterized by a linear counting rate of $2 \times 10^5$ photons/s/pixel, a dynamic range of 32 bits, and a pixel size of $130 \times 130$ μm$^2$. Their threshold can be tuned over an energy range of 4–35 keV. The pixel number and the active area differs from one detector to the other (see Figure 3).

All the active surfaces of the detectors are made of similar modules. The D5 has eight more modules compared to the S70, and is stacked in a column to increase the detection area with an inter module gap of 5 mm for electronic wirings and connections. Both detectors have a frame rate of 100 Hz. Whereas, the WOS, with a higher frame rate of 250 Hz, has 10 modules mounted as shown in Figure 3c (5 rows × 2 columns) in order to maximize the detection area, particularly for WAXS measurements. The horizontal gap between the two modules of the second row is higher than the others to give way for an opening 10 mm in diameter in the active area. On the backside of the WOS detector, a 10 cm exit opening increases the exit angular range. Even though it can be used on the diffractometer to have a large angular range, when installed on the SAXS granite table the WOS can detect WAXS signals and let the transmitted beam go through its opening. The transmitted beam, thus, gets detected by the D5 mounted one to three meters downstream of the sample. Accordingly, this configuration allows for performing simultaneous SAXS/WAXS experiments (Figure 4).

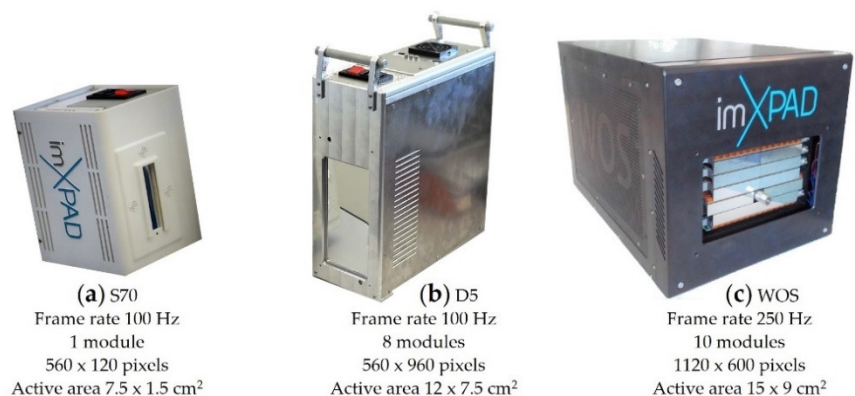

**Figure 3.** The ImXpad detectors available at the D2AM beamline: (**a**) the S70, (**b**) the D5, and (**c**) the WOS (wide-angle X-ray scattering (WAXS) Open for small-angle X-ray scattering (SAXS)).

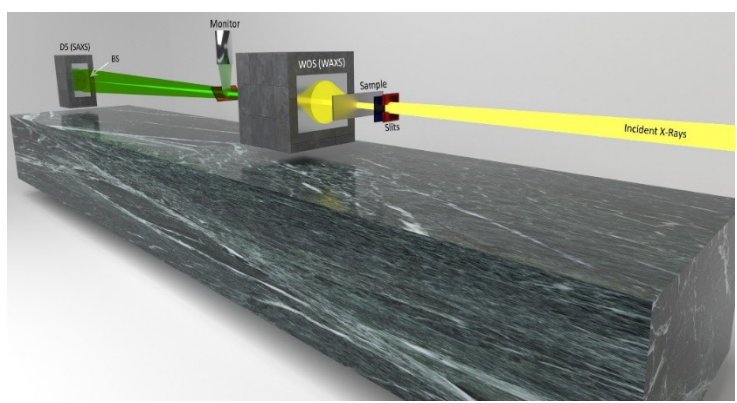

**Figure 4.** Experimental setup of a simultaneous SAXS/WAXS experiment on the SAXS bench of the D2AM beamline. The X-ray beam is focused on the beam stop (BS) positioned in front of the D5 detector. Scatter-less slits are usually mounted 10 cm away from the sample. A photomultiplier is used as a monitor with a retractable kapton foil.

These detectors may be fixed on the granite table on motorized *z–y* translation stages or on the detector arm of the diffractometer that can rotate around the *z*- and *y*-axes. The sample–detector distance may consequently range between 20 to 5 m or 10 to 2 m for SAXS and XRD measurements, respectively. The weight load and space limitations on the granite table, together with the high weight load capacity and the kappa geometry of the diffractometer, makes D2AM a well-suited beamline for synchrotron in situ and operando X-ray diffraction measurements. Besides the sample environment equipment that can be used from the ESRF instrument pool (heating coils, traction machine, etc.), a variety of sample environments were developed on the beamline with collaborators and scientists to meet specific requirements and needs of the beamline user community.

*2.3. Sample Environments*

A wide variety of sample environments can be installed on the D2AM beamline and benefit from the X-ray scattering advanced characterization techniques for in situ measurements. While beamline users can bring their own equipment, we will limit our discussion in this paper to the most frequently used ones that are always available at the beamline.

2.3.1. "QMAX" Very High Temperature Furnace

For achieving high temperatures, a new furnace (Figure 5) capable of heating samples under primary vacuum, air, or controlled atmosphere up to 1700 °C was developed at the beamline in collaboration with beamline users [12]. It can be mounted on the goniometer head on the diffractometer, allowing to conduct in situ wide-angle or small-angle measurements under conventional out-of-plane, in-plane, or grazing incidence. The "QMAX furnace" consists of a heating metallic resistor supported by a ceramic plate, 20 mm in diameter, which connects the sample thermally to the heating element. Two domes (Be and PEEK) were specifically made for this furnace. These domes diffract at different angles when crossed by an X-ray beam. Choosing one or the other is based on the required temperature/gas environment, as well as the region of the reciprocal space that scientists would like to explore to prevent parasitic scattering/overlapping of sample-dome signals. Two controlled gas lines were designed and calibrated in terms of flow rate for oxygen and nitrogen. The heating plate and the thermocouples are read and controlled by a Nanodac controller from Eurotherm by Schneider Electric. The furnace can handle a heating rate of up to 50 °C/s with a thermal stability of ±1 °C. According to these characteristics, this furnace is very well-suited for the study of a large number of scientific topics related to the behavior of metals at high temperature (solid–gas reactivity, thermal expansion, phase transitions, phase separations, self-ordering processes, etc.).

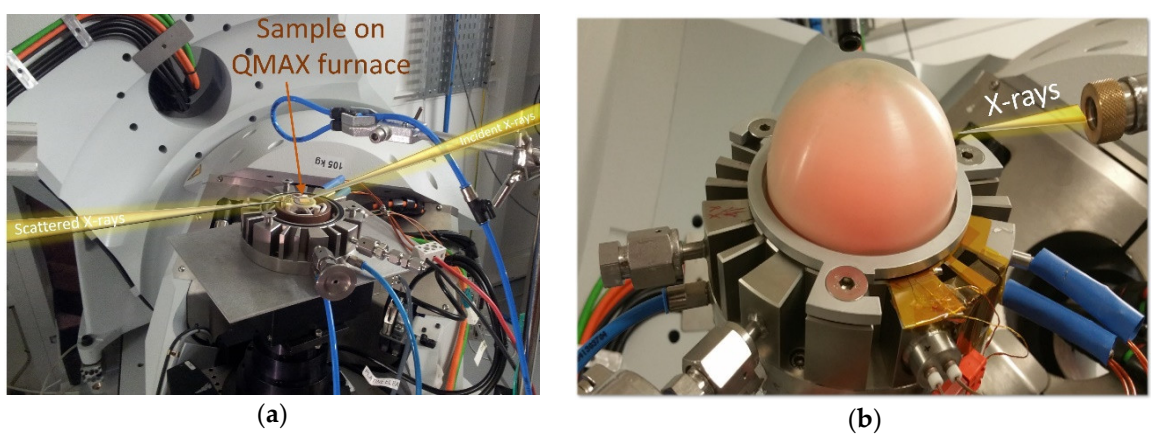

(**a**)　　(**b**)

**Figure 5.** The QMAX heating stage with a sample fixed on a ceramic plate heated (**a**) in air and (**b**) under a PEEK (PolyEther Ether Ketone) dome.

2.3.2. Small-Angle X-ray Scattering (SAXS) Furnace

Another furnace, developed in-house together with SIMaP (Science et Ingénierie des Matériaux et Procédés) colleagues, is dedicated to SAXS and WAXS measurements. It is installed on a translational motorized stage inside a vacuum vessel. It is designed to work in a transmission geometry where the beam goes through the sample. Two configurations are available with different opening diameters for scanning larger samples. While the first one (Figure 6) with an aperture of 4 mm can heat up to 900 °C, the second one, with an opening of 25 mm, can heat up to 500 °C. Controlled by a Nanodac station, these furnaces are used for SAXS and WAXS in situ measurements. De Geuser et al. [18] investigated, using in situ SAXS, the relationship between alloy processing and its chemical properties by studying aging effects on nanostructured precipitations in samples having a composition gradient. Using a similar experimental approach on cold-pressed as-milled powders, Deschamps et al. were able to investigate the precipitation kinetics in as-milled oxide dispersion-strengthened steel powders during heating [19]. Currently, a new SAXS furnace is under development to help achieve temperatures up to 1200 °C.

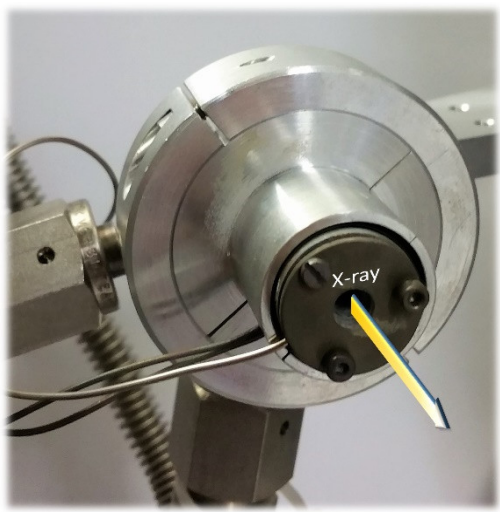

**Figure 6.** SAXS furnace with a 4 mm opening.

2.3.3. Cryostat

A cryostat can be mounted on the kappa diffractometer (Figure 7) for low-temperature studies, typically 10 K is achieved under a secondary vacuum using two Be domes. A He gas compressor (ARScryo, Advanced Research Systems, Macungie, PA, USA) is used to cool down the sample where the He is circulated in a closed circuit. A heater head, set below the sample stage, helps to not only regulate the temperature but also to heat up to 800 K. The heating is controlled by a Lakeshore 336 temperature controller (Lake Shore Cryotronics Inc., Westerville, OH, USA). Even though many cables and tubes have to be connected to the cryostat, all the diffractometer translations and rotations are maintained, giving access to all of the reciprocal space and to conduct in situ measurements without instrumental limitations. Grenier et al. observed the Verwey transition in a magnetite ultrathin film grown on Ag that was recently detected, near 120 K, by conducting in situ resonant X-ray scattering [20]. A low-temperature phase transition in a periodic approximant to an icosahedral quasicrystal were investigated by Yamada et al. This is an order–disorder transition accompanied by a lattice distortion from cubic to monoclinic at low T (190 K). The cryostat allowed not only the study of phase transition but also of the pre-transitional fluctuations and their divergence when approaching Tc from high temperatures [21,22].

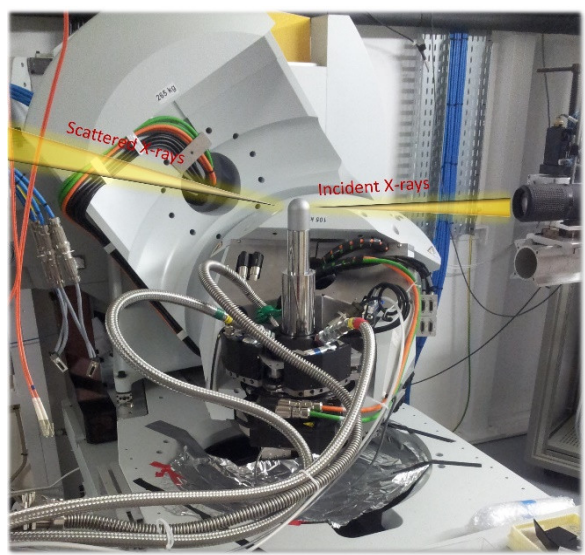

**Figure 7.** The D2AM cryostat mounted on the beamline kappa diffractometer.

Even though the mentioned sample environments are the most requested and frequently used at the beamline, other equipment for specific in situ characterization has been used, or can be easily accommodated. For instance, a tensile testing machine (Deben UK Ltd., Suffolk, UK) was installed on the D2AM diffractometer to study the effects of cyclic deformations on the evolution of dislocation structures as a function of grain size and film thickness [23]. New fabrication processes, e.g., additive manufacturing, may also benefit from the in situ X-ray characterization tool. It is now possible, for example, to map the microstructural evolution during the melting and the solidification processes in selective laser melting (3D printing) using X-ray diffraction. A thorough investigation of these processes will help optimize the printing settings for more reliably produced parts. In addition, a hot isostatic pressing (HIP) device may be designed and mounted on the beamline to perform temporally and spatially resolved X-ray measurements in situ. They may also be designed to look into mechanisms involved in processes for improving mechanical properties with sintering powder or achieving a diffusion weld.

*2.4. Data Treatment and Analysis*

The multidimensional data sets recorded at the beamline are treated directly after each measurement, allowing one to drive the experiment on the fly. Python-based short scripts were written by beamline scientists to perform specific treatments, such as the conversion from angular to reciprocal space, azimuthal integrations and correction of distortions induced by the detectors' geometries based on the PyFAI (Fast Azimuthal Integration using Python) suite [24], data visualization (3D maps, cuts), and fittings. Due to the specificity and complexity of the conducted experiments, these scripts undergo continuous minor or major modifications depending on the experimental configuration and the information that needs to be accessed. The Jupyter Notebooks platform is used to keep track and document the data treatment process. The beamline is, thus, building a modular python-based software that can be adapted to each experiment, creating a data treatment electronic logbook that helps to process the data in future similar experiments. With the variety of experimental configurations and experiments that are being conducted at the beamline, this modular approach appears to be best suited for performing fast and efficient online data analysis and to meet the challenges of big data.

**3. Conclusions**

The compactness and high efficiency of in situ devices, combined with the high load capacity of the D2AM precise kappa diffractometer and the three fast photon-counting detectors, make the beamline a very powerful tool for advanced, in situ, structural, and non-destructive characterization. This approach will provide more insight into the kinetics of specific phenomena such as phase transitions, precipitations, oxidation, corrosion, diffusion, and doping. This approach will contribute to understanding the structural evolution under internal and external, varying conditions and reactions, such as new alloys and new fabrication technologies (e.g., additive manufacturing) and how they may be optimized and designed. Even though the available technology offers great opportunities for the fine investigation of matter, a concurrent synchrotron source upgrade (EBS) is taking place to bring the most brilliant synchrotron X-ray source. With the EBS program, the D2AM beamline will deliver micrometer beam-spot size for in situ micro structural imaging and higher resolution mappings. In addition, an increase in X-ray beam coherence will open a completely new realm of possibilities in terms of imaging, both spatially and temporally, the structural parameters in materials. New devices will be acquired and developed to take advantage of the new source characteristics. At the D2AM beamline, new KB mirrors with low surface roughness, new monochromator designs for collecting higher intensities in the focused beam, and new photon counting detectors are under development and will equip the beamline after 2020 for faster experiments and higher resolution mapping.

**Author Contributions:** Conceptualization, G.A.C., Na.B., Ni.B. and S.A.; Investigation, G.A.C., Na.B., F.D.G., R.G. and Ni.B; Resources, G.A.C., Na.B., Ni.B. and S.A.; Writing—Original Draft Preparation, G.A.C.; Writing—Review & Editing, R.G., Na.B., N.B. and S.A.; Visualization, G.A.C.; Project Administration, Na.B.

**Funding:** Part of the development of the beamline was made within the framework of the QMAX Project No. ANR-09-NANO-031, funded by the French National Agency (ANR) within its Nanosciences, Nanotechnologies, and Nanosystems program (P3N2009), as well as the EQUIPEX ANR-11-EQPX-0010 'CRG/F' funded by the French National Agency (ANR). As part of the French CRG (F-CRG) beamlines, the operating budget of the D2AM beamline is funded by a French CEA-CNRS consortium.

**Acknowledgments:** The beamline development would not have been possible without the support of the French Consortium CEA-CNRS that we thank. The success of the beamline in this field is a result of many contributions and commitments of users and scientists who were closely involved in the beamline investigations. Thanks to their efforts and to their feedback, the beamline continues to evolve to offer what is best for the user community.

**Conflicts of Interest**: The authors declare no conflict of interest. The funders had no role in the design of the study; in the collection, analyses, or interpretation of data; in the writing of the manuscript, or in the decision to publish the results.

**References and Notes**

1. Filippelli, E.; Chahine, G.; Borbely, A. Evaluation of intragranular train and average dislocation density in single grains of a polycrystal using K-map scanning. *J. Appl. Cryst.* **2016**, *49*, 1814–1817, doi:10.1107/S1600576716013224.
2. Deschamps, A.; Bastow, T.J.; de Geuser, F.; Hill, A.J.; Hutchinson, C.R. In situ evaluation of the microstructure evolution during rapid hardening of an Al-2.5Cu-1.5Mg (wt. %) alloy. *Acta Mater.* **2011**, *59*, 2918–2927, doi:10.1016/j.actamat.2011.01.027.
3. De Geuser, F.; Malard, B.; Deschamps, A. Microstructure mapping of a friction stir welded AA2050 Al-Li-Cu in the T8 state. *Philos. Mag.* **2014**, *94*, 1451–1462, doi:10.1080/14786435.2014.887862.
4. Matringe, C. Nanostructuration Bidimensionnelle de Surfaces Vicinales de Saphir: Etude Quantitative Par Diffusion et Diffraction des Rayons x sur Sources de Lumière Synchrotron. Ph.D. Thesis, Université de Limoges, Limoges, France, 2016.
5. Couturier, L.; De Geuser, F.; Deschamps, A. Determination of the volume fraction of precipitates in a nitride Fe-0.354 wt.% C-2.93 wt.% Cr model alloy by anomalous small angle X-ray scattering. *Mater. Charact.* **2018**, *135*, 134–138, doi:10.1016/j.matchar.2017.11.036.
6. Couturier, L.; De Geuser, F.; Deschamps, A. Direct comparison of Fe-Cr unmixing characterization by atom probe tomography and small angle scattering. *Mater. Charact.* **2016**, *121*, 61–67, doi:10.1016/j.matchar.2016.09.028.
7. Revenant, C.; Benwadih, M.; Maret, M. Self-organized nanoclusters in solution-processed mesoporous In-Ga-Zn-O thin films. *Chem. Commu.* **2015**, *51*, 1218–1221, doi:10.1039/c4cc08521c.
8. EQUIPEX ANR-11-EQPX-0010 'CRG/F' funded by the French National Agency (ANR).
9. Hazemann, J.L.; Nayouf, K.; de Bergevin, F. Modelisation by finite elements of sagittal focusing. *Nucl. Instrum. Methods Phys. Res. Sect. B* **1995**, *97*, 547–550, doi:10.1016/0168-583X(94)00731-4.
10. Ferrer, J.L.; Simon, J.P.; Bérar, J.F.; Caillot, B.; Fanchon, E.; Kaikati, O.; Arnaud, S.; Guidotti, M.; Pirocchi, M.; Roth, M. D2AM, a beamline with high-intensity point-focusing fixed-exit monochromator for multiwavelength anomalous diffraction experiments. *J. Synchrotron Radiat.* **1998**, *5*, 1346–1356, doi:10.1107/S0909049598004257.
11. Andrault, D.; Barrett, R.; Bayat, S.; Berkvens, P.; Biasci, J.-C.; Billinge, S.; Boulanger, B.; Bouteille, J.-F.; Bravin, A.; Brun, E.; Carla, F.; et al. *The ESRF orange book: ESRF Upgrade Programme Phase II (2015-2022): Technical Design Study*; ESRF: Grenoble, France, 2014.
12. QMAX Project (Quantitative analysis of the microstructured thin films. High-resolution X-ray diffraction and grazing incidence small angle X-ray scattering coupling) No. ANR-09-NANO-031-03 funded by the French National Agency (ANR) in the frame of its program in Nanosciences, Nanotechnologies and Nanosystems (P3N2009).
13. Boulle, A.; Masson, O.; Guinebretière, R.; Dauger, A. Miscut angles measurement and precise sample positioning with a four-circle diffractometer. *Appl. Surf. Sci.* **2001**, *180*, 322–327, doi:10.1016/S0169-4332(01)00369-5.

14. Garcia, J.; Subias, G.; Blasco, J.; Sánchez, M.; Beutier, G. Resonant X-ray scattering study of charge superstructures in layered La 2−*x* Ca *x* CoO 4 ± δ (0.4 ≤ *x* ≤ 0.7) and La 1.5 Sr 0.5 CoO 4 compounds. *Phys. Rev. B* **2018**, *97*, 085111, doi:10.1103/PhysRevB.97.085111.
15. Thune, E.; Fakih, A.; Matringe, C.; Babonneau, D.; Guinebretière, R. Understanding of one dimensional ordering mechanisms at the (001) sapphire vicinal surface. *J. Appl. Phys.* **2017**, *121*, 015301, doi:10.1063/1.4973341.
16. Matringe, C.; Fakih, A.; Thune, E.; Babonneau, D.; Arnaud, S.; Blanc, N.; Boudet, N.; Guinebretière, R. Symmetric faceting of a sapphire vicinal surface revealed by Grazing Incidence Small-Angle X-ray Scattering 3D mapping. *Appl. Phys. Lett.* **2017**, *111*, 031601, doi:10.1063/1.4985339.
17. Bérar, J.F.; Boudet, N.; Breugnon, P.; Caillot, B.; Chantepie, B.; Clemens, J.C.; Delpierre, P.; Dinkespiler, B.; Godiot, S.; Meessen, C.; et al. XPAD3 hybrid pixel detector applications. *Nucl. Instrum. Methods Phys. Res. Sect. A* **2009**, *607*, 233–235, doi:10.1016/j.nima.2009.03.208.
18. De Geuser, F.; Styles, M.J.; Hutchinson, C.R.; Deschamps, A. High-throughput in-situ characterization and modeling of precipitation kinetics in compositionally graded alloys. *Acta Mater.* **2015**, *101*, 1–9, doi:10.1016/j.actamat.2015.08.061.
19. Deschamps, A.; De Geuser, F.; Malaplate, J.; Sornin, D. When do oxide precipitates form during consolidation of oxide dispersion strengthened steels? *J. Nucl. Mater.* **2016**, *482*, 83–87, doi:10.1016/j.jnucmt.2016.doi:10.017.
20. Grenier, S.; Bailly, A.; Ramos, A.Y.; De Santis, M.; Joly, Y.; Lorenzo, J.E.; Garaudée, S.; Frericks, M.; Arnaud, S.; Blanc, S.; et al. Verwey transition in a magnetite ultrathin film by resonant X-ray scattering. *Phys. Rev. B* **2018**, *97*, 104403, doi:10.1103/PhysRevB.97.104403.
21. Yamada, T.; Euchner, H.; Gómez, C.P.; Takakura, H.; Tamura, R.; de Boissieu, M. Short- and long-range ordering during the phase transition of the Zn6Sc 1/1 cubic approximant. *J. Phys. Condens. Matter* **2013**, *25*, 205405, doi:10.1088/0953-8984/25/20/205405.
22. Euchner, H.; Yamada, T.; Schober, H.; Rols, S.; Mihalkovic, M.; Tamura, R.; Ishimasa, T.; de Boissieu, M. Ordering and dynamics of the central tetrahedron in the 1/1 Zn6Sc periodic approximant to quasicrystal. *J. Phys. Condens. Matter* **2012**, *24*, 415403, doi:10.1088/0953-8984/24/41/415403.
23. Renault, P.O.; Sadat, T.; Godard, P.; He, W.; Guerin, P.; Geandier, G.; Blanc, N.; Boudet, N.; Goudeau, P. Continuous cyclic deformations of a Ni/W film studied by synchrotron X-ray diffraction. *Surf. Coat. Technol.* **2017**, *332*, 351–357, doi:10.1016/j.surfcoat.2017.06.082.
24. Ashiotis, G.; Deschildre, A.; Nawaz, Z.; Wright, J.P.; Karkoulis, D.; Picca, F.E.; Kieffer, J. The fast-azimuthal integration Python library: pyFAI. *J. Appl. Cryst.* **2015**, *48*, 510–519, doi:10.1107/S1600576715004306.